# Classification of colorectal primer carcinoma from normal colon with mid-infrared spectra

B. Borkovits[1]  |  E. Kontsek[2]  |  A. Pesti[2]  |  P. Gordon[3]  |  S. Gergely[4]  |
I. Csabai[5]  |  A. Kiss[2]  |  P. Pollner[1,6]

[1]Department of Biological Physics, Eötvös Loránd University, Budapest, Hungary

[2]Department of Pathology, Forensic and Insurance Medicine, Semmelweis University, Budapest, Hungary

[3]Department of Electronics Technology, Budapest University of Technology and Economics, Budapest, Hungary

[4]Department of Applied Biotechnology and Food Science, Budapest University of Technology and Economics, Budapest, Hungary

[5]Department of Physics of Complex Systems, Eötvös Loránd University, Budapest, Hungary

[6]Data-Driven Health Division of National Laboratory for Health Security, Health Services Management Training Centre, Semmelweis University, Budapest, Hungary

**Correspondence**
E. Kontsek, Department of Pathology, Forensic and Insurance Medicine, Semmelweis University, Budapest 1091, Hungary.
Email: kontsek.endre@semmelweis.hu

**Funding information**
National Research, Development and Innovation Office (Hungary), Grant/Award Numbers: K_18 128881, OTKA K128780; Data-Driven Health Division of National Laboratory for Health Security, Grant/Award Number: RRF-2.3.1-21-2022-00006; MILAB Artificial Intelligence National Laboratory, Grant/Award Number: RRF-2.3.1-21-2022-00004; European Union, Grant/Award Numbers: Nr. 10109571, RRF-2.3.1-21-2022-00015; Ministry for Culture and Innovation from the source of the National Research, Development and Innovation Fund., Grant/Award Number: DKOP-23 Doctoral Excellence Program

**Abstract**

In this project, we used formalin-fixed paraffin-embedded (FFPE) tissue samples to measure thousands of spectra per tissue core with Fourier transform mid-infrared spectroscopy using an FT-IR imaging system. These cores varied between normal colon (NC) and colorectal primer carcinoma (CRC) tissues. We created a database to manage all the multivariate data obtained from the measurements. Then, we applied classifier algorithms to identify the tissue based on its yielded spectra. For classification, we used the random forest, a support vector machine, XGBoost, and linear discriminant analysis methods, as well as three deep neural networks. We compared two data manipulation techniques using these models and then applied filtering. In the end, we compared model performances via the sum of ranking differences (SRD).

**Abbreviations:** 1Conv, convolutional network with 1 Conv2D layer; 1DUNet, U-Net architecture; 3Dense, MLP with 3 dense layers; CRC, colorectal primer carcinoma; FT-IR, Fourier transform infrared spectroscopy; LDA, linear discriminant analysis; MLP, multilayer perceptron; NC, normal colon; PCA, principal component analysis; RFC, random forest classifier; ROC curve, receiver operating characteristic curve; SRD, sum of ranking differences; SVC, support vector classifier; *t*-SNE, *t*-distributed stochastic neighbor embedding; TMA, tissue microarray; XGBC, XGBoost classifier.

## 1 | INTRODUCTION

Cancer is one of the main causes of death in the developed world.[1] Among the various types worldwide, colorectal cancer is among the most common according to various statistics.[2] In many cases, treating it requires surgical removal of cancerous tissue.

Generally, there are five stages of colorectal cancer numbered from 0 to 4. The first stage is called carcinoma, which is simply cancerous cells that are located only in the mucosa. In this stage, the cancer only affects the inner layer of the colon. Then, in the next two stages, the cancer gradually spreads to the deeper layer of the colon. In the final two stages, the cancer spreads to lymph nodes and causes metastases in other organs. Treatment-wise, the last two stages prove to be the originators of most problems as the cancer cannot be isolated to the colon anymore.[3]

When it comes to treatment, the procedure will depend on the overall health and age of the patient as well as the stage their cancer is in. For patients in stage 0, surgical resection is the stand-alone method for treatment. Then, for those in stage 1, surgical resection of the malignant polyp or partial colectomy (removal of the colon) of tumor and local lymph nodes is. Meanwhile for stage 2, surgical resection or if that was not successful, chemotherapy (for instance, 5-FU or leucovorin) might become necessary. Then, for Stage 3, surgery followed by adjuvant chemotherapy, including FOLFOX (combined chemotherapy). In addition, for some advanced colon cancers in Stage 3 therapy may include surgery, neoadjuvant (pre-surgery) chemotherapy, and radiation, while for people who are not healthy enough for surgery, radiation therapy and or chemotherapy. Finally, for stage 4, the following procedures are usually utilized: radiotherapy; chemotherapy with, for example, FOLFIRI or FOLFOX (combined chemotherapies), immunotherapy (pembrolizumab or nivolumab), targeted therapies, palliative surgery or stenting, radiofrequency ablation, and radioembolization.[4]

Regarding surgical resection, finding the border that separates normal and altered tissue is not necessarily trivial. To remedy this, multiple different techniques are used for identification during surgery. The applications of techniques based on spectroscopy not in the visible light spectrum are being studied by many research groups to positively influence the detection of lesions and treatment of colorectal cancer. Mid-infrared (MIR) spectroscopy is a non-invasive and non-destructive technique; therefore, it could be used during surgery without potential risks to the patient and has definite benefits over techniques such as ionizing radiation. Furthermore, it might also provide a substantial amount of chemical information to detect differences between normal and cancerous tissue.[5] This property of the technique makes it ideal to couple with machine learning to effectively separate these tissues and lesions. In addition, tissue microarrays (TMA) provide a suitable method to collect and contain numerous tissue samples. These samples are placed in a single paraffin block with the block then being customized in size and shape to better suit the planned analysis, thus creating the needed tissue cores. Finally, the cores get placed onto glass slides and then passed for analysis.[6]

While working with spectroscopic and imaging methods, we can distinguish between labeled and unlabeled approaches. In the case of labeled approach, we pick out fluorescent dye (e.g., indocyanine green or ICG short) whose presence we monitor in different areas. With this approach, it is common to measure a very small spectral window as it decreases the time of measurement substantially. On the other hand, with the unlabeled approach, we examine a larger range of the spectrum and then, based on the spectral fingerprint, make conclusions about the presence of the respective components in the given areas. Typically, this technique comes with a longer measuring time.

We opted to use Fourier Transform Infrared Spectroscopy (FT-IR) as our technique for spectral analysis as it is a widely accepted method for identifying different types of cancer.[7–10] Furthermore, it has already been applied to examine samples that contain tissue with colorectal cancer.[11,12] In addition, as part of an experiment in the literature, the FT-IR spectrometer was coupled with an endoscope to record spectra measured during endoscopy, and those spectra were then processed through a classification algorithm to determine, whether the patients had cancer or not. They managed to obtain quite accurate results with this setup.[13] Meanwhile, FT-IR has many other applications as well. For instance, it can be used for the classification of different bacteria.[14] Apart from medical analysis, the usage of this technique also saw some success when applied to food analysis, such as measurement of edible oils, milk, butter, and so forth.[15,16]

General FT-IR instrumentation works in the following way. First, the source emits an infrared beam through the aperture that manages the magnitude of the incident energy that will reach the sample. Then, the interferometer applies on to the beam a method called spectral encoding that yields its output as an interferogram signal. After that, the signal exits the interferometer and reaches the sample. The next part of the process depends on what analysis we wish to conduct. We can choose reflection or transmission. Respectively, the beam will be reflected off or transmitted through the surface of the sample. Then, the beam reaches the detector, where the interferogram signal will be measured. Finally, a digitized signal is sent to a computer that is wired to the setup. Here, the signal is Fourier transformed to yield the desired infrared spectra.[17,18] These spectra are obtained in the form of a matrix, in which a row represents a spectrum, while the columns represent the wavenumbers at which the measurement of absorbance took place. On the other hand, if we are dealing with hyperspectral images, the desired results are obtained in the form of a three-dimensional spectral hypercube. The first two dimensions are spatial and contain the $x$ and $y$ coordinates of the pixel at which a spectrum was measured. The third dimension is the spectral one, containing the obtained spectra.[19]

In this project, we first applied to the already preprocessed dataset two dimensionality reduction algorithms: principal component analysis (PCA) and $t$-distributed stochastic neighbor embedding ($t$-SNE). These tools were utilized to visualize the difference between tissue and background data in two respective latent spaces where one data point represented one spectrum. PCA is a linear method that generates new features as the linear combinations of original ones,[20] while $t$-SNE is a non-linear technique that maps points into the new space via their pairwise similarity defined with the aid of a Student's $t$ distribution centered on one of the points in the original space.[21] We applied both of these techniques because they generated significantly distinct latent representations, aiming to discern any consequential impact on our results.

Afterwards, a number of supervised machine learning methods were utilized to make classification of tissue cores of normal colon (NC) and colorectal primer carcinoma (CRC) types. A family of these models consists of the ensemble methods that use results provided by multiple independent models to yield their output. The ensemble methods can be further divided into subcategories, like bagging (bootstrap aggregating) and boosting. A bagging method that we used is the random forest classifier (RFC), which yields the average of predictions made by multiple decision trees; hence, it is an ensemble method.[22]

We utilized a method from the boosting family as well, the XGBoost Classifier (XGBC) algorithm. It is based on decision trees too; however, instead of averaging over results obtained from many of them, it strives to improve or rather "boost" the accuracy of predictions made by decision trees using their previously weaker results. This method uses gradient tree boosting.[23]

In addition, we utilized the support vector classifier (SVC) with its radial basis function kernel.[24] Then, we also applied linear discriminant analysis (LDA) that makes classification by fitting Gaussian densities to each class.[25]

Besides the above-mentioned methods, we also considered deep neural networks. These neural networks see wide use in tasks such as medical image segmentation and signal reconstruction and many more.[26] They offer many customization options such as adding multiple different layers and choosing from a wide variety of loss functions and optimizers to further increase the complexity of the model while trying to decrease the run time and increase the efficiency of the predictions.

A well-known deep neural network architecture for processing biomedical image data is U-Net.[27] Certain versions of it contain different optimization techniques, such as the inclusion of even more layers or the manipulation of a few parameters. A few examples of its many versions are U-Net$^{++}$, U-Net3D, Adversarial U-Net, and Residual U-Net with each of them modified to better suit the purpose of their creators.[28]

A well-trained machine learning model has the capability to make good predictions in very short time, therefore having advantage when used during surgery in real time. In this project, we collected spectral data obtained from tissue cores via MIR spectroscopy to serve as the input for several machine learning models and then compared the efficiency of these models.

## 2 | MATERIALS AND METHODS

### 2.1 | Sample preparation

To obtain a mid-infrared reflective surface, glass slides were covered with thin-film metal layers by vacuum evaporation. In a high-vacuum chamber, where the glass slides were secured onto the rotary sample holder, an electron-beam was then applied. For 20 min at $10^{-4}$ Pa at an accelerating voltage of 7 kV and beam current of 200 mA, the aluminum was evaporated to reach a layer thickness of 150 nm.[29]

The tissue samples used in the analysis were prepared at Semmelweis University, Department of Pathology, Forensic and Insurance Medicine. The use of human FFPE samples was approved by the Hungarian Medical Research Council, Budapest, Hungary (no. 61303-2/2018/EKU and 32191-2/2019/EKU). After extracting tissue from the patients, the samples went through the following process.[30] First, the samples were fixed using buffered formalin of 10% at room temperature for 24 h. Then, they were dehydrated in a six part series of graded ethanol (first: 50%, second: 75%, third 10991128, 0, Downloaded from https://analyticalsciencejournals.onlinelibrary.wiley.com/doi/10.1002/cem.3542 by Cochrane Hungary, Wiley Online Library on [22/03/2024]. See the Terms and Conditions (https://onlinelibrary.wiley.com/terms-and-conditions) on Wiley Online Library for rules of use; OA articles are governed by the applicable Creative Commons License and fourth: 96%, fifth and sixth: 100%) for 6 × 10 min, and in xylene for 2 × 15 min inside a tissue processor. At 56 °C temperature paraffin embedding was carried out. Two hours later, the samples were put into a new paraffin block. Tissues belonging to NC and CRC representative areas were cored and extracted from the donor blocks to then be put into the recipient paraffin blocks that were prepared using a TMA Master II (3DHistech Ltd, Budapest, Hungary) device. Each new block received two cores from each tissue. Afterwards, the TMA blocks were incubated for 30 min at a temperature of 37 °C to allow the process of adhesion between cores and paraffin. Next, sections of 2 μm thickness were cut for H&E staining and then infrared imaging. The process of H&E consisted of the following steps: All sections were stained with hematoxylin and with eosin for 10 and 1 min, respectively. As for infrared imaging, additional 2 μm thick slides were cut and then doused in xylene two times for 10 min for deparaffining. Two example tissue cores can be seen in Figure 1. We performed data acquisition from 62 cores that were taken from 17 patients: from 11 patients 4 cores/patient, from 6 patients 3 cores/patient. It should be noted that out of the 4 cores of a patient, 2 were NC and 2 were CRC cores with all patients suffering from colorectal cancer. For 3 cores/patient, the cores were either 1 NC and 2 CRC, or 2 NC and 1 CRC.

## 2.2 | Spectral acquisition

To collect the desired spectra by measuring our chosen tissue samples, we used a Spotlight 400 FT-IR Imaging System[31] with a customized setup so that it obtained hyperspectral images in the mid-infrared range. The Spotlight 400 microscope (Perkin Elmer Inc., Waltham, Massachusetts, USA) was connected to Spectrum 400 spectrophotometer used for scanning images. The instrument is shown in Figure 2 and is located at Budapest University of Technology and Economics, Department of Applied Biotechnology and Food Science. The setup and its SpectrumIMAGE R1.6.5.0396 software (Perkin Elmer Inc., Waltham, Massachusetts, USA) offered several parameters to better tune the analysis. We could choose between reflection and transmission modes, this time opting for the prior, as the tissue cores occupied the majority of the area of measurement and were non-transparent. It was also possible to limit the wavenumber interval at which we wished to obtain our results, effectively manipulating the length of the spectra.

Furthermore, we were able to set the spectral resolution, which meant increasing (higher resolution) or decreasing (lower resolution) the difference between two wavenumbers. At lower resolution, one tended to obtain noisier and harder to interpret data. On the other hand, a higher-than-needed resolution setting decreased the number of data points of the spectra so much that the measurement might have omitted useful information. Finally, we could alter the speed at which the interferometer was taking its measurements along with the coaddition number.

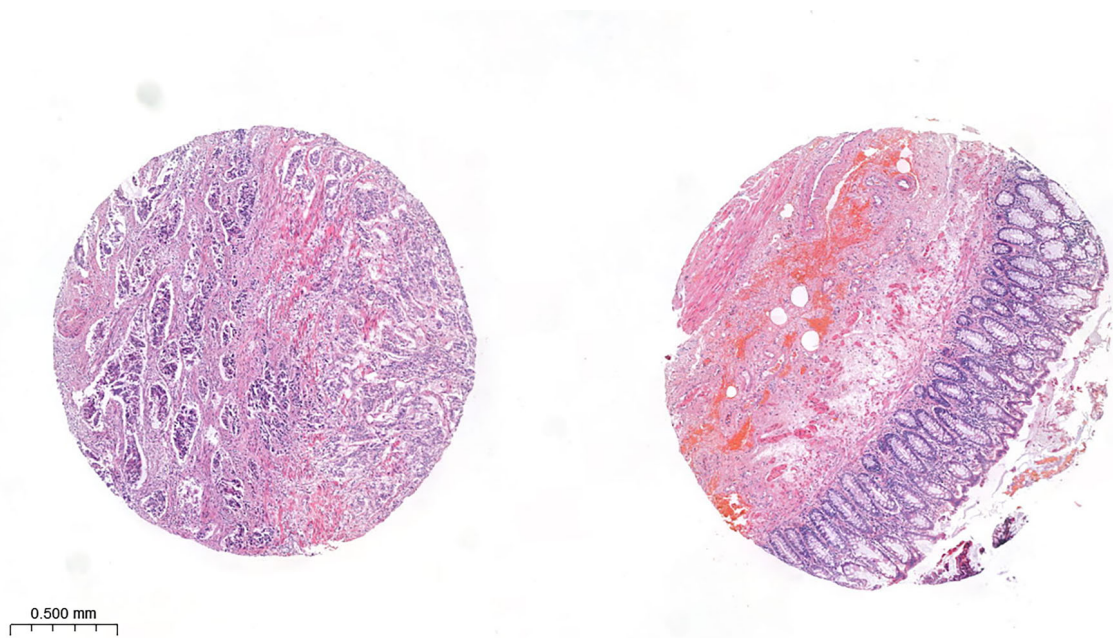

**FIGURE 1** Two bright field images taken from a TMA during deparaffining, where the colors signify the hematoxylin–eosin staining. On the left, there is a CRC, while on the right, a NC tissue core. The SlideViewer software was used to display these images.

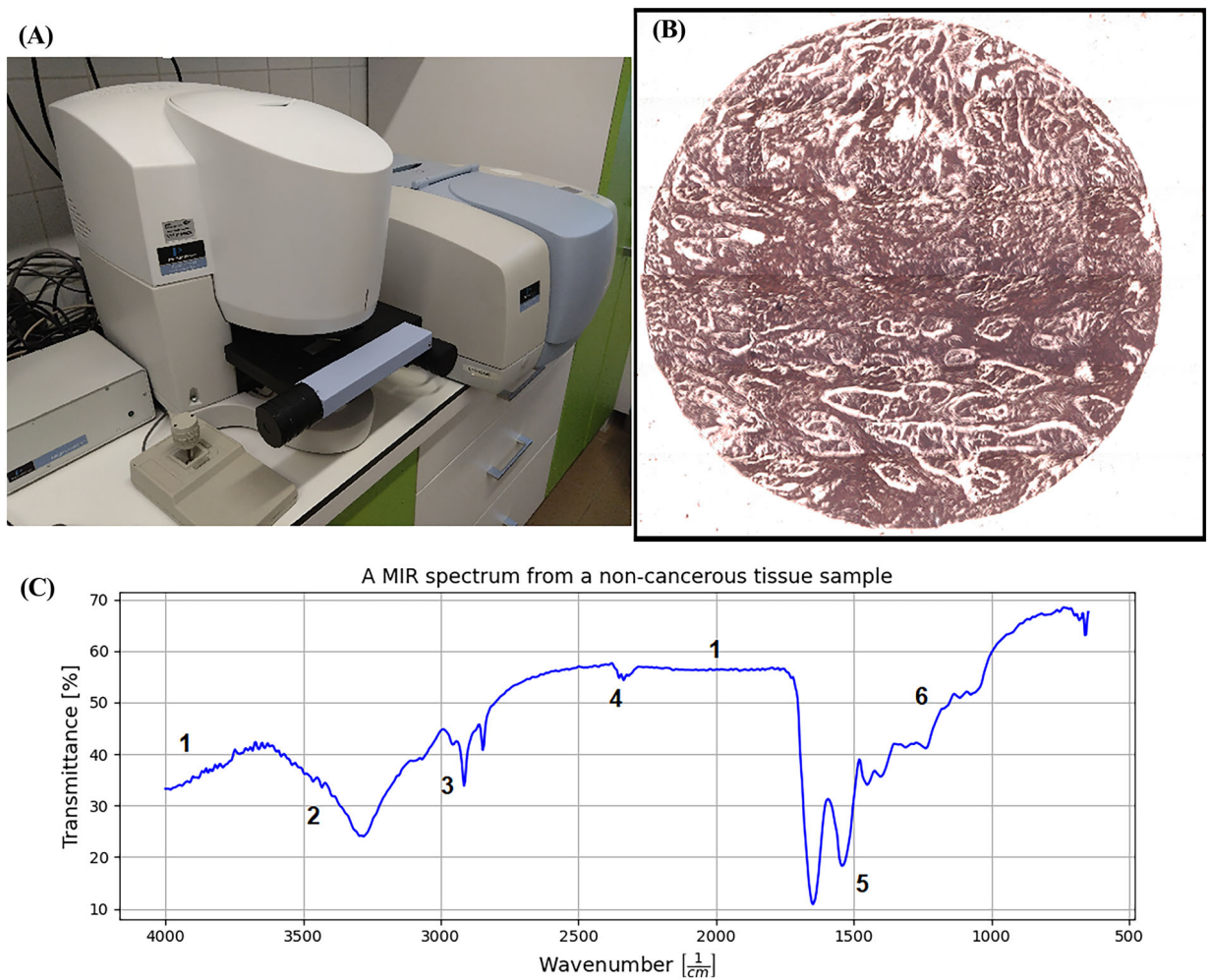

**FIGURE 2** (A) A photo of the Spotlight 400 FT-IR Imaging System we chose to work with. On the right is the source of the infrared beam, while on the left is the interferometer coupled with the sampling compartment and the detector. (B) Depicting the area of measurement inside the black outline. The area of this square is $2200 \times 2200$ μm$^2$, though it may seem a bit distorted because of image conversion. (C) An example of obtaining chemical information from a mid-IR spectrum. The numbers signify the existence of vibrations caused by the following substances: 1. water vapor, 2. nitrogen-hydrogen bond, 3. hydrocarbon, 4. carbon dioxide, 5. protein, and 6. carbohydrates. The x-axis follows the convention as ordering the spectrum by increasing wavelength but using wavenumbers for labeling.

After inserting the slide into the sample compartment of the instrument and moving the interferometer to focus on a chosen core, we could finally set our parameters for the measurement. Firstly, we chose the area of the square within which we wished to conduct the measurement. The diameter of a tissue core with a more regular shape was roughly 2200 μm; thus, one such square would look like the one showcased in Figure 2. For the rest of the parameters, we determined a good setup by conducting calibrating measurements that focused on minimizing measuring time and noise level. Finally, we settled for 8 cm$^{-1}$ as the spectral resolution, 16 for coaddition (scans per pixel), 2.2 $\frac{cm}{s}$ for interferometer speed, and $25 \times 25$ μm$^2$ as pixel size. With these settings, the measurement of one core took roughly 30 min. As we took our measurements in the 4000–748 cm$^{-1}$ wavelength range, we obtained one spectral value for every fourth wavelength (4000, 3996, 3992, …), yielding a total of 814 points per spectra coupled with the coordinates of the pixel at which the spectra were collected. This data structure was then stored in an FSM Perkin Elmer Spotlight IR binary file (.fsm) that contained the data matrix of the coupled spatial and spectral data alongside an array that was made up of the wavelengths. A final column was then added that contained an identifier of what type of core (NC or CRC) a given spectrum had been taken from. In the end, we obtained 7744 spectra per tissue core.

## 2.3 | Data manipulation

The total number of spectra we received from the measurements: $62 \times 7744 = 480{,}128$. However, this includes spectra taken from the background of the slides as well. First, we converted the spectral values in transmittance percentages to absorbance using the following formula:

$$A = \log_{10}\left(\frac{1}{\frac{T[\%]}{100}}\right) = 2 - \log_{10}(T[\%]), \tag{1}$$

where $A$ is absorbance and $T$ means transmittance.

Afterwards, we needed to decide how to handle background data. We applied two methods to separate spectra taken from tissue and from background. The first method is the "slicing operation." Here, the exclusion of every spectrum that was measured from a pixel outside of the coordinate ranges $-750 < X < 750$ and $-750 < Y < 750$, because this area roughly contained the largest inner rectangle of most circular tissue cores. After performing this, there remained 223,200 spectra. Should we achieve good results for the machine learning models with the usage of this method, we could reduce the area of measurement and therefore the measurement time significantly. However, as it can be viewed on Figure 3, a good portion of the tissue was also omitted from further analysis, and therefore, meaningful information might have been lost.

To address the shortcomings of the previous method, we also applied $K$-means clustering with two clusters meant to denote tissue and background data. $K$-means clustering is an unsupervised algorithm; thus, it performs classification without the need of labels. It first randomly initializes a given number of centers. In our case, that number was 2. Then, it assigns points to their closest center while it also refreshes the center by moving it to the average of the points that belong to it. We applied this method for every single tissue core. In the end, we received two clusters as it is depicted in Figure 3. Afterwards, we proceeded to mark every spectrum with its given $K$-means label, which was 0 if the spectrum belonged to the tissue cluster or 1 if it belonged to the background cluster. Then, we kept only those spectra that belonged to the tissue class (263,278 spectra). This number was a bit larger than in the case of the previous method; however, these spectra contained more tissue data and significantly less background data because $K$-means clustering was capable of filtering out the background near the middle of the core as well.

Finally, to prepare the spectra for analysis using machine learning algorithms, we applied a method called standard normal variate that normalized a spectrum by subtracting its mean and dividing by its standard derivation. After applying this to every spectrum, the mean and standard deviation of each became 0 and 1, respectively. The aim of this technique was to make all spectra have comparable absorbance levels, thus efficiently applying corrections for optical path length and light scattering changes.[32,33]

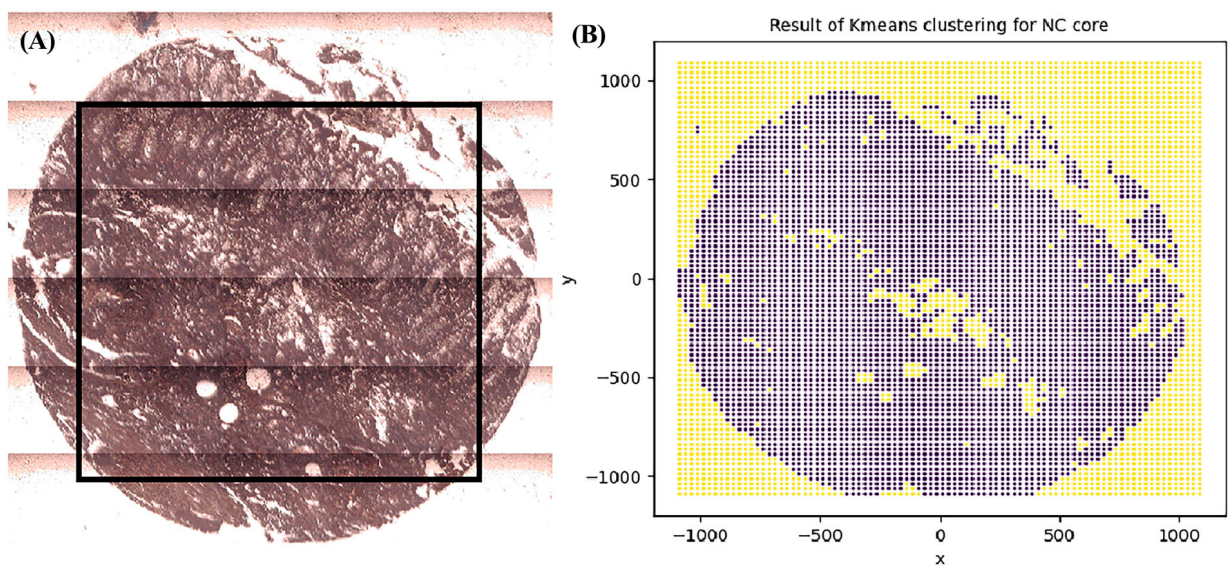

**FIGURE 3** (A) This bright field image was recorded by the microscope. The area within the black square was kept as the result of the slicing operation. The loss of some tissue data is visible around the desired surface. (B) The result of the $K$-means clustering for tissue and background data. The black area was kept.

As explained earlier, a spectrum contained 814 values, meaning 814 features for the supervised models. Then, we used two blanking filters directly after obtaining the spectra. One of those was a $CO_2$ blanking filter that served the purpose of removing the influence of carbon dioxide from the spectrum. This is handled by excluding all data between 2390 and 2280 cm$^{-1}$ wavenumbers. The other was a $H_2O$ blanking filter that served a similar purpose by excluding data between 4000 and 3500 cm$^{-1}$, as well as those between 1900 and 1300 cm$^{-1}$ wavelengths.[34] By applying these two filters, from the 814 features, there remained 512 that we then inputted into the supervised models.

## 2.4 | Methods used for analysis

For our first step to compare the slicing operation and the K-means clustering, we applied PCA and t-SNE. These methods were used to showcase the effect the two operations had on a tissue core and whether they had been successful in separating tissue and background data. However, later on, we did not utilize the dimensionality reduction offered by PCA and t-SNE but opted to input the original data to the supervised models.

Before applying the supervised methods, we did the train-test split in the following way: Out of the 62 tissue cores, 58 landed in the training set, while 4 was kept in the test set. In the training set, there were 30 NC and 28 CRC cores; meanwhile, there were 2 NC and 2 CRC cores in the test set. After separating the sets, we shuffled the spectra in each. It should be noted that shuffling the data inside the set before or after the splitting is required in order to prevent the model from learning distinct patterns of the training set that stemmed only from the order of the data. Here, we shuffled after splitting to avoid overlaps. For validation, we used 12 such different train-test splits and then averaged the yielded model efficiencies and gave their standard deviation as the error term.

For our version of the U-Net deep neural network (1DUNet), the input dimension parameter was the number of features, namely, the number of wavenumbers at which we obtained our spectral values during measurement (814 or 512). We built a sequential model using the Tensorflow.Keras API in python. Because our spectra were one-dimensional as we had not trained on or predicted the spatial coordinates, we needed to adjust the layers accordingly. Instead of Conv2D, Maxpooling2D, and Conv2DTranspose (deconvolutional) layers, we implemented their one-dimensional versions and switched their kernel size to a single number signifying the fact that we were dealing with one-dimensional data. Max pooling creates a pooled feature map with the maximum values of the patches or local clusters of neurons. We also kept the dropout and concatenate layers in our implementation. We also included an UpSampling1D layer after each deconvolution that resized the output of the deconvolutional layer to matching dimensions. Finally, to make the model yield the desired results, we added a dense layer with a given activation function as the output layer with 2 neurons denoting the number of classes we wanted to predict. We opted to use the sigmoid activation function. As we were trying to solve a binary classification problem, we used binary cross-entropy as loss function. We also inputted the adaptive momentum estimation (Adam) optimizer.

In addition to our U-Net implementation, we will also be using a multilayer perceptron with 3 Dense (fully connected) layers, a simpler convolutional network with a Conv2D (two-dimensional convolution), a Maxpooling2D (two-dimensional max pooling), and two dense layers. These networks were denoted respectively as 3Dense and 1Conv.

All in all, seven different supervised techniques were used to make predictions: RFC, XGBC, LDA, SVC, and the 3 neural networks. At this point, it should be noted that with the exception of the neural networks, all unsupervised and supervised methods were coded in python using the scikit-learn library.[35]

The efficiency of the methods was measured with a receiver operating characteristic (ROC) curve and a confusion matrix,[36] and we show four aggregated quantities that are derived from these two complex descriptors: the $AUC$ (area under ROC curve) score and[37]

$$\text{Sensitivity} = \frac{TP}{TP+FN}, \text{Specificity} = \frac{TN}{TN+FP}, ACC = \frac{TP+TN}{TP+TN+FP+FN}, \qquad (2)$$

where their scales range from 0 to 1 with 1 being a perfect score, but a model that is better than a random predictor has $AUC > 0.5$.

We used the sum of ranking differences (SRD) to compare our methods regarding the similar their *ACC* scores. The evaluation scheme of the SRD can be viewed in the article published by Bajusz et al[38] in Figure 1, which we do not include to avoid copyright violation. We chose this ranking method because it disregards the linear or non-linear nature of models and treats them completely independently. The model that had the smallest SRD, even if not ideal, was decidedly the best method. Because we had seven supervised models to compare, identifying the superior one was achievable.[39] It should be noted that some models may have the same SRD, in which case we are talking about degeneracy, meaning that these models are indistinguishable. We created a matrix from the *ACC* scores of each model including the results of validations yielded from the three data manipulation methods: the slicing operation, *K*-means clustering, and the atmospheric correction. Because the higher the *ACC*, the better the model, we set the maximum value in each row in the matrix as reference point. We also generated random numbers and computed their SRD values to then compare them with those of the models. This acted as validation to see whether a model behaved randomly or in a deterministic way. The farther away the actual SRD values were from the random ones, the less the models would be associated with random behavior.[39]

## 3 | RESULTS

### 3.1 | Comparison of the two data manipulation methods

In this subsection, we examined which previously defined data separating technique proved superior. Of course, the two methods we compared were the slicing operation and *K*-means clustering.

Firstly, we applied the two operations to a singular tissue core. In both cases, we rendered color labels to the spectra measured from that core. A spectrum received "blue" for color if it was to be omitted from further analysis. In the case of the slicing operation, this meant all data outside of the rectangle, where $-750 < X < 750$ and $-750 < Y < 750$. For the clustering, blue denoted every spectrum that was put into the background data cluster. On the other hand, spectra that received the color "red" were to be kept as input for the supervised learning models, meaning they were measured inside the earlier mentioned rectangle, or simply were classified as members of the tissue data cluster.

Then, we applied PCA and *t*-SNE to visualize the two sets of points per method for a tissue core. The results can be viewed in Figure 4. Both subfigures showcased that the new structure in the latent space agreed with the results of *K*-means clustering better than with those of the slicing operation. On the left side of the figure, it was shown that a not so negligible amount of sliced off data mingled with the kept spectra. Because this was the case both with PCA and *t*-SNE, as well as the phenomenon not being present in the *K*-means clustering visualization, it was safe to say that this particular sliced off data denoted tissue data specifically. This aligned with the initial statement that by executing the slicing operation, we cut off a moderate portion of our tissue data. Whether we had lost a significant amount of information or not was for the supervised learning algorithms to determine. The results of this comparison were showcased in Tables 1 and 2. The *AUC* score of the SVC method is missing because of the algorithm overflowing the memory when trying to retain the class probabilities for further calculation.

It should be noted that no hyperparameter tuning had been done as of writing this article; therefore, all models used default parameters. First things to note were the relatively high error bars for every single model and for each efficiency metric. These numbers were this high despite averaging the results of 12 different train-test splits, meaning that some of the test sets and thus tissue cores proved notably different to predict.[40] Another reason for this could have been an imbalance in the number of spectra from the two classes; however, we circumvented this issue by making sure that all training and test datasets were set up to be balanced in that regard. Because of that, we did not use SMOTE for oversampling because for the original number of 480,128 spectra, we would only have gained an additional 15,488 new spectra meaning a mere 3% increase in data size. Finally, this behavior could have been caused by training sets not having the appropriate size to provide stable predictions. This problem could easily be remedied in the future, as we had already been planning to conduct more measurements on new tissue cores.

On the other hand, the high amount of uncertainty could also be contributed to the fact that the size of the test set was relatively small compared with the general test dataset size used in supervised analysis. Out of the four cores included in a test set if we had misclassified only one, it would have increased the uncertainty in efficiency significantly. We defined such test sets because during pathological analysis, we expected to receive only the small number of tissue cores for identification and thus validated model performance accordingly.

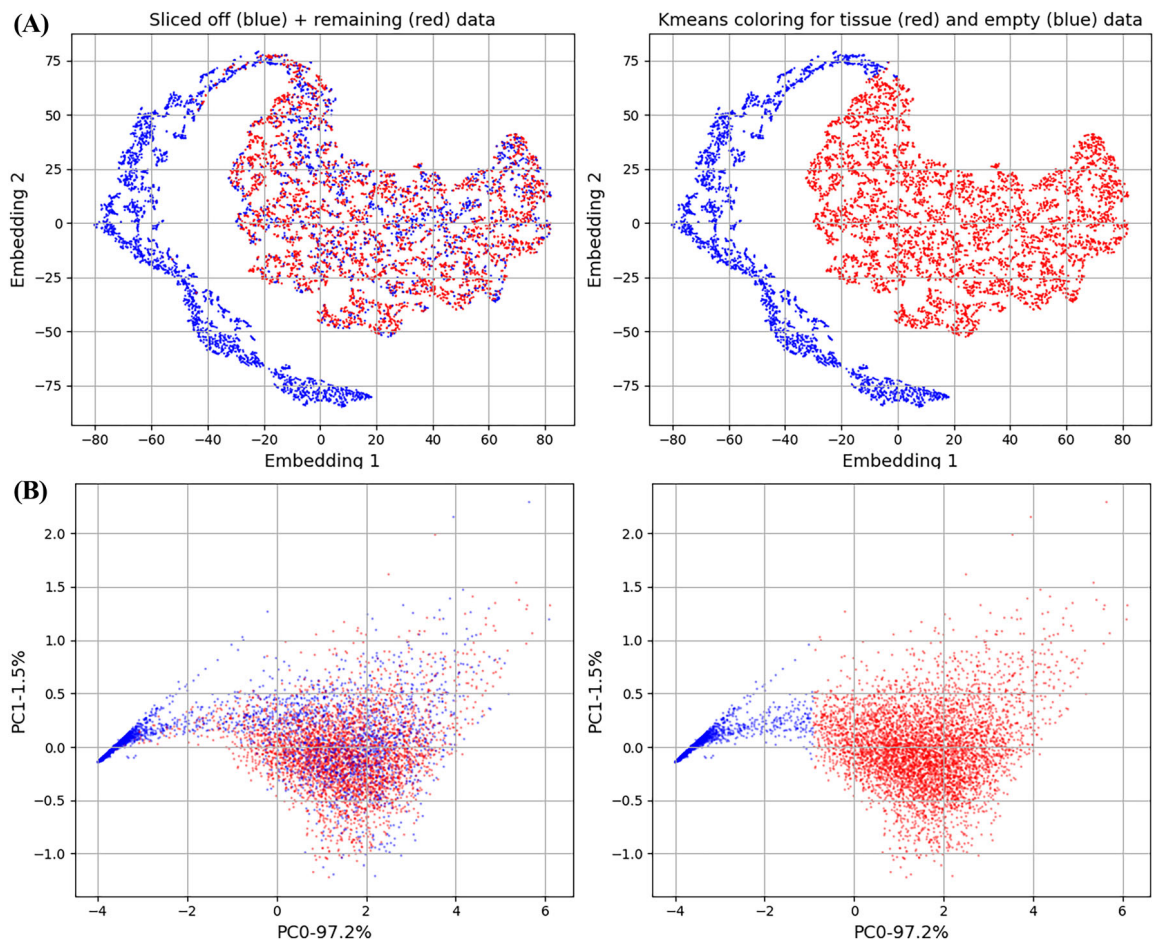

**FIGURE 4** Latent space visualization of a tissue core. In both subfigures the left image corresponds to the slicing operation while the right to K-means clustering. The colors represent data that we used in further analysis (red) and data that were left out from it (blue). One point belonged to one spectrum. (A) t-SNE visualization. (B) PCA visualization.

**TABLE 1** The efficiency metrics of the machine learning models when their input was the data that remained after performing the slicing operation.

| Methods | ACC | AUC | Sensitivity | Specificity |
| --- | --- | --- | --- | --- |
| Random forest classifier | 0.73 ± 0.17 | 0.79 ± 0.19 | 0.70 ± 0.23 | 0.76 ± 0.21 |
| XGBoost classifier | 0.75 ± 0.15 | 0.82 ± 0.15 | 0.74 ± 0.20 | 0.77 ± 0.20 |
| Linear discriminant analysis | 0.79 ± 0.09 | 0.88 ± 0.08 | 0.80 ± 0.16 | 0.77 ± 0.20 |
| Support vector classifier | 0.77 ± 0.10 | - | 0.78 ± 0.16 | 0.77 ± 0.20 |
| MLP with 3 dense layers | 0.78 ± 0.18 | 0.82 ± 0.19[a] | 0.76 ± 0.27[a] | 0.81 ± 0.27[a] |
| 2D convolutional neural network | 0.78 ± 0.21 | 0.81 ± 0.22[a] | 0.79 ± 0.31[a] | 0.78 ± 0.29[a] |
| 1D U-Net neural network | 0.81 ± 0.17 | 0.83 ± 0.18[a] | 0.76 ± 0.27[a] | 0.86 ± 0.18[a] |

*Note*: The errors were obtained by taking the standard deviation from running the models on 12 different train-test split sets, while the values themselves are averages obtained the same way.
Abbreviations: ACC, overall accuracy; AUC, area under ROC curve score; MLP, multilayer perceptron.
[a] Of course, the efficiency metrics cannot achieve a score higher than 1; these occurred only because of our use of standard deviation as error bars.

Meanwhile, the average overall accuracy for the models in both cases stood between 75% and 85%. When it comes to diagnosing cancer, these numbers are simply too low not to mention that other studies dealing with colorectal cancer data showed a stable higher than 90% accuracy for as the result of classification, albeit with different machine learning

**TABLE 2** The efficiency metrics of the machine learning models when their input was the data that remained after performing $K$-means clustering.

| Methods | ACC | AUC | Sensitivity | Specificity |
| --- | --- | --- | --- | --- |
| Random forest classifier | $0.75 \pm 0.16$ | $0.84 \pm 0.17$[a] | $0.76 \pm 0.25$[a] | $0.76 \pm 0.19$ |
| XGBoost classifier | $0.77 \pm 0.14$ | $0.87 \pm 0.12$ | $0.77 \pm 0.22$ | $0.79 \pm 0.17$ |
| Linear discriminant analysis | $0.82 \pm 0.18$ | $0.86 \pm 0.26$[a] | $0.82 \pm 0.26$[a] | $0.79 \pm 0.21$ |
| Support vector classifier | $0.80 \pm 0.11$ | - | $0.80 \pm 0.17$ | $0.79 \pm 0.21$ |
| MLP with 3 dense layers | $0.79 \pm 0.18$ | $0.83 \pm 0.17$ | $0.80 \pm 0.23$[a] | $0.79 \pm 0.29$[a] |
| 2D convolutional neural network | $0.81 \pm 0.17$ | $0.84 \pm 0.18$[a] | $0.78 \pm 0.28$[a] | $0.85 \pm 0.21$[a] |
| 1D U-Net neural network | $0.82 \pm 0.20$[a] | $0.84 \pm 0.21$[a] | $0.78 \pm 0.29$[a] | $0.87 \pm 0.17$[a] |

*Note*: The errors were obtained by taking the standard deviation from running the models on 12 different train-test split sets, while the values themselves are averages obtained the same way.

Abbreviations: ACC, overall accuracy; AUC, area under ROC curve score; MLP, multilayer perceptron.

[a] Of course, the efficiency metrics cannot achieve a score higher than 1; these occurred only because of our use of standard deviation as error bars.

methods.[41] Meanwhile, projects revolving around FT-IR techniques and spectrum level classification via traditional models or modified neural networks managed to reach similar or higher efficiency scores as well.[42–47] However, in some cases, these studies put a lot more focus on the preprocessing of the spectra or even using the different derivatives as input for the machine learning models, which could definitely be a viable option for us to better our results. The accuracy can be improved by adding new slides to the cohort and increasing the number of spectra.

A more notable result could be seen in the *AUC* column of both tables. Not a single result at the lowermost point of their error bar touches the region beneath or at 50%, meaning that the two classes were differentiated correctly every single time. In addition, the average values for sensitivity and specificity were very close to each other; thus, it is safe to say that overfitting had been done for neither of the two classes.

Unlike in other studies[48,49] where a spectrum represented a single cancerous cell, here, a spectrum was a representation of a larger tissue part, which contained numerous different singe-cells and the models still labeled most spectra correctly. Further studies examined the efficiency of similar models by applying unsupervised pattern recognition or clustering on tissue cores on which the location of areas denoting benign, malignant, and so on parts was already known.[50–52] Compared with them, we had no prior knowledge of such areas in any tissue core, meaning that even though we might have had vastly different spectra based on those areas, we still managed to predict the correct label for most of them.

To continue, we had to decide which of the two data manipulation methods should be kept. Despite the almost identical results yielded for the model efficiencies, we have chosen $K$-means clustering because of its ability to filter out background data from inside the area of the tissue cores, as well to its capability to retain tissue spectra that otherwise would be cut from the analysis by the slicing operation. This was proven by describing the results depicted in Figure 4.

## 3.2 | Filtering

For our next step, we performed filtering on the dataset. As mentioned before, a spectrum was made out of 814 points, each taken at a different wavenumber. Then, we applied the two previously described blanking filters to reduce the number of points to 512 and then once again checked the results of the machine learning models that are displayed in Table 3. In case of the classic models (RFC, XGBC, LDA, and SVC), the results changed by a negligible amount. The average values increased or decreased by a few percentages, the most notable decrease (12%) present at the specificity of the SVC model, but this was the only one exception. Generally, however, this barely changing behavior proved to be a benefit, as the models gave almost the same results for 512 features as they did for 814, signifying a meaningful dimension reduction and thus decreased run times for the models without the decline of their efficiencies.

Meanwhile, the *ACC*, sensitivity and specificity averages of the neural networks remained somewhat the same as well, however, their *AUC* scores increased uniformly. In addition, the final and probably the most meaningful result that came from applying the filters was the consistent shrinking of the error bars regarding the three neural networks. This behavior shows the difference between the 12 predictions decrease in a uniform manner, meaning the data that we filtered

**TABLE 3** The efficiency metrics of the machine learning models when their input was the data that remained after performing *K*-means clustering and then we applied the blanking filters.

| Methods | *ACC* | *AUC* | Sensitivity | Specificity |
| --- | --- | --- | --- | --- |
| Random forest classifier | 0.76 ± 0.09 | 0.84 ± 0.10 | 0.81 ± 0.10 | 0.72 ± 0.18 |
| XGBoost Classifier | 0.76 ± 0.11 | 0.84 ± 0.13 | 0.80 ± 0.11 | 0.72 ± 0.19 |
| Linear Discriminant Analysis | 0.75 ± 0.16 | 0.80 ± 0.23[a] | 0.74 ± 0.24 | 0.74 ± 0.18 |
| Support Vector Classifier | 0.72 ± 0.13 | - | 0.75 ± 0.16 | 0.67 ± 0.21 |
| MLP with 3 Dense layers | 0.81 ± 0.10 | 0.88 ± 0.10 | 0.81 ± 0.17 | 0.81 ± 0.19 |
| 2D convolutional neural network | 0.83 ± 0.14 | 0.89 ± 0.13[a] | 0.79 ± 0.21 | 0.88 ± 0.14[a] |
| 1D U-Net neural network | 0.81 ± 0.11 | 0.89 ± 0.10 | 0.79 ± 0.19 | 0.83 ± 0.14 |

*Note*: The errors were obtained by taking the standard deviation from running the models on 12 different train-test split sets, while the values themselves are averages obtained the same way.
Abbreviations: ACC, overall accuracy; AUC, area under ROC curve score; MLP, multilayer perceptron.
[a] Of course, the efficiency metrics cannot achieve a score higher than 1; these occurred only because of our use of standard deviation as error bars.

out had a definite effect in worsening the predictions. The shrinking of the error bars occurred in the case of the RFC models with the exception of its specificity. We have already mentioned the decrease in the specificity of the SVC method; therefore, later, we would need to pay special attention to certain specificity results to account for unique behavior.

To sum up, we compared our slicing operation and *K*-means clustering to separate tissue and background data. Although the supervised machine learning models yielded practically the same results for both of these options, we found that *K*-means clustering managed to exclude background data from the inner parts of tissue cores while also keeping tissue spectra that were cut off by the slicing operation. In addition, we applied our two blanking filters and learned that using them came with multiple benefits, making them a viable first step in dimensional reduction.

## 3.3 | Sum of ranking differences

As the final part of our project, we tried to determine which model proved to be superior at differentiating spectra from the two defined classes via the SRD method. The to be processed matrix contained all *ACC* scores yielded by the 12 - different runs for both the slicing operation, *K*-means clustering, and the filtering. This meant 36 points per method. Although fewer points could have given conceptually the same results, with this many, showcasing and visualizing the ranking of the models turned out to be more clear. As our SRD implementation, we used a Python program made by Dávid Bajusz.[53] The ranking of supervised learning models is displayed in Figure 5.

First of all, it seemed that results for all models differ eloquently from those of the random numbers; therefore, neither model behaves randomly. Secondly, a clear ranking was displayed. Once again, it should be noted that given a fewer number of points in the matrix, we could have seen degeneracies, where the models with the same SRD would have been indistinguishable. However, clusters could still somewhat be seen on the figure. These were the pairs of 1DUNet and 1Conv, the pair of 3Dense and SVC, and finally the pair of XGBC and LDA methods. To verify this, we made use of the *k*-fold cross-validation technique built into the Python implementation and plotted the results as a box plot complete with outlier detection. After applying 10-fold cross-validation, we observed our U-Net network proving superior to the smaller convolutional network nine out of ten times.

As an additional way to examine the supposed clusters in Figure 5, the Wilcoxon signed-rank test was performed where the null hypothesis stood for the *ACC* scores of two methods having been drawn from the same distribution.[54] In Table 4, we can see that the null hypothesis could only be discarded for the SVC—3Dense pair, because its $p_{value}$ turned out to be smaller than 0.05. This means that the difference between the efficiency of the two methods was statistically significant. In the other two cases, the same differences were statistically not significant, and thus, without the already performed cross-validation, we would not have been able to sufficiently differentiate between the elements of those pairs.

It should not have been a surprise that the XGBC outperformed the RFC model, as it is basically an upgraded version of decision tree models and not just an average of them. Meanwhile, it turned out that all three neural networks outperformed all classic machine learning models with the most complex one, namely, the 1DUNet, proving itself as the best method. On the other hand, this neural network with its 21.60% SRD was still far from an ideal model with 0%

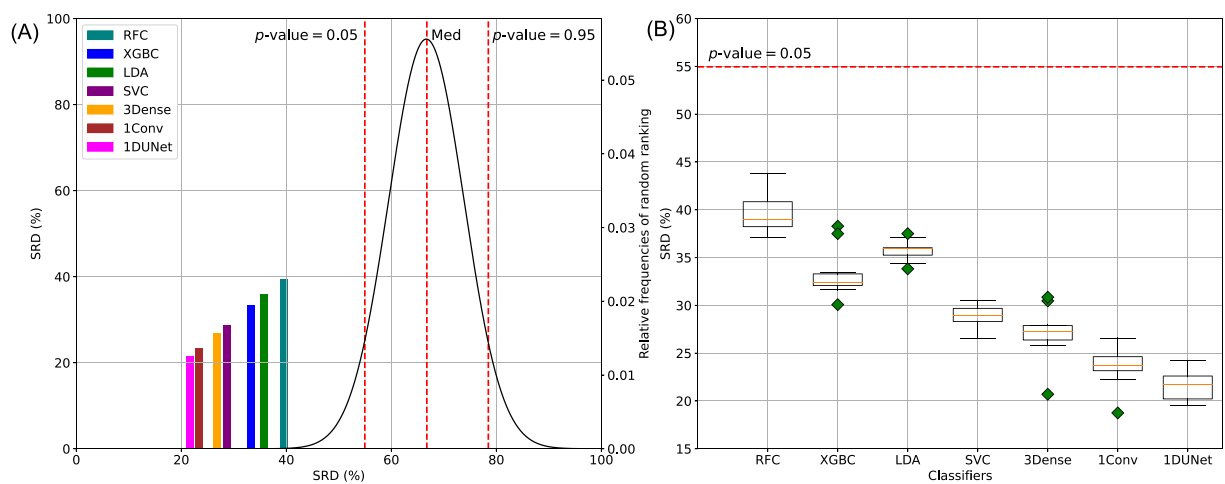

**FIGURE 5** (A) Visualization of the sum of ranking differences-based ranking of the models. The black curve represents the SRDs of the random numbers generated during the comparison. Every model stands apart from the random distribution, meaning that each model behaved deterministically. (B) Ten-fold cross-validation was performed to determine the exact ranking of our models. The green markers denote outliers and the yellow lines showcase the median of each box.

**TABLE 4** Wilcoxon signed-rank test performed on the pairs of methods that seemed to belong in the same cluster in Figure 5.

| Pairs of methods | $p_{value}$ |
| --- | --- |
| XGBC—LDA | 0.52 |
| SVC—3Dense | 0.03 |
| 1Conv—1DUNet | 0.81 |

SRD. As we planned to increase the size of our dataset, it would be advisable to repeat the ranking process later and examine any change, as well as to utilize a different ranking tool such as analysis of variance (ANOVA) to validate these particular results.

## 4 | CONCLUSIONS

In this project, we measured mid-infrared spectra from the surface of pathological glass slides containing tissue cores of normal colon and colon with colorectal primer carcinoma via FT-IR technique. Then, the gained spectra were collected to make a database and to serve as input for machine learning models to differentiate between the two tissue classes. Before the analysis, removing the majority of background data was required; thus, a comparison between our slicing operation and $K$-means clustering was made to determine which fulfilled the task better. It was shown that although the overall results were eerily similar, results provided by the clustering were more inline with those obtained from the latent space comparison of the two methods. Therefore, we applied blanking filters for dimensional reduction on data that remained after performing $K$-means clustering. This filtering, while retaining the previous efficiency scores, turned out to also provide many smaller benefits, such as the overall reduction of the difference between predictions made during validation. Because of these, we were successful in reducing the size of our input dataset by a bit more than one third. In addition, using the sum of ranking differences with cross-validation, where the multi-criteria consisted of the overall accuracy of the classifiers gained after using the slicing operation, spectrum selection with $K$-means clustering and after applying the blanking filters, we determined that our one-dimensional implementation of the U-Net deep neural network architecture was our best classifier.

In the future, we will increase the size of our database by conducting measurements on more pathological slides and thus increase the size of the training set as well. To achieve higher efficiency, we will proceed with hyperparameter tuning for the supervised learning models, while also trying out different sizes for the test set. As for the SRD method, as said before, we plan to utilize ANOVA to create a more comprehensive ranking of the models. Our hope for the future firstly is to publish our database and for it to prove useful in pathological work and secondly to develop a method to classify cancerous and non-cancerous tissue during surgery via mid-infrared spectroscopy.

## ACKNOWLEDGEMENTS

We would like to offer our gratitude to János Slezsák, Károly Héberger, Fanni Csiza, and Erzsébet Kovács for the theoretical help they provided as well as to József Stéger and Dávid Visontai for their technical help with handling the database.

## PEER REVIEW

The peer review history for this article is available at https://www.webofscience.com/api/gateway/wos/peer-review/10.1002/cem.3542.

## DATA AVAILABILITY STATEMENT

Research data are not shared.

## ORCID

*B. Borkovits*  https://orcid.org/0009-0004-0696-7586
*E. Kontsek*  https://orcid.org/0000-0002-8098-1392
*A. Pesti*  https://orcid.org/0000-0001-6706-6221
*S. Gergely*  https://orcid.org/0000-0003-1945-526X
*I. Csabai*  https://orcid.org/0000-0001-9232-9898
*A. Kiss*  https://orcid.org/0000-0002-7453-3163
*P. Pollner*  https://orcid.org/0000-0003-0464-4893

**How to cite this article:** Borkovits B, Kontsek E, Pesti A, et al. Classification of colorectal primer carcinoma from normal colon with mid-infrared spectra. *Journal of Chemometrics*. 2024;e3542. doi:10.1002/cem.3542